# Decoherence processes during active manipulation of excitonic qubits in semiconductor quantum dots

Q. Q. Wang,[1,2,3] A. Muller,[1] P. Bianucci,[1] E. Rossi,[1] Q. K. Xue,[2] T. Takagahara,[4]

C. Piermarocchi,[5] A. H. MacDonald,[1] and C. K. Shih[1*]

[1]*Department of Physics, University of Texas at Austin, Austin, Texas 78712*

[2]*International Center for Quantum Structures (ICQS), Institute of Physics, The Chinese Academy of Sciences, Beijing 100080, P. R. China*

[3]*Department of Physics, Wuhan University, Wuhan 430072, P. R. China*

[4]*Department of Electronics and Information Science, Kyoto Institute of Technology, Kyoto 606-8585, Japan*

[5]*Department of Physics and Astronomy, Michigan State University, East Lansing, Michigan 48824*

(Submitted to PRL)

Using photoluminescence spectroscopy, we have investigated the nature of Rabi oscillation damping during active manipulation of excitonic qubits in self-assembled quantum dots. Rabi oscillations were recorded by varying the pulse amplitude for fixed pulse durations between 4 ps and 10 ps. Up to 5 periods are visible, making it possible to quantify the excitation dependent damping. We find that this damping is more pronounced for shorter pulse widths and show that its origin is the non-resonant excitation of carriers in the wetting layer, most likely involving bound-to-continuum and continuum-to-bound transitions.

The current topic of quantum computation presents a wide range of challenges to physical science [1], particularly the search for candidates for solid-state quantum bits (qubits). Semiconductor quantum dots (QDs) are attractive because they possess energy structures and coherent optical properties similar to, and dipole moments larger than, those of atoms [2,3]. Efforts in the past few years have led to successful observations of Rabi oscillations (ROs) of excitonic states [4-9], the hallmark for active manipulation of qubits in QDs. However, all found that ROs damped out very quickly when the external field is increased. Because QDs contain a macroscopic number of atoms, this strong decoherence process must be due to unwanted coupling to other degrees of freedom. Identification of the underlying mechanism is difficult precisely because of this macroscopic nature. Yet such understanding plays the most crucial role in future development of quantum information technology in semiconductors. Through manipulations of high quality factor excitonic qubits in InGaAs QDs, we have studied the underlying mechanism for decoherence processes during active manipulation. More specifically, we have found that this strong decoherence process is manifested through indirect excitations of carriers in the wetting layer whose composition is highly fluctuating.

We study $In_{0.5}Ga_{0.5}As$ self-assembled QD (SAQD) samples grown by molecular beam epitaxy (MBE). The details of growth processes are given in [10]. These QDs are embedded in a GaAs matrix with a wetting layer of roughly 5 monolayers thickness. The dots have an average lateral size, height, and dot-to-dot distance of 20-40 nm, 4.5 nm and 100 nm, respectively, characterized using cross-sectional scanning tunnelling microscopy. There are three excitonic levels involved: The exciton vacuum (labelled as $|0\rangle$) when there is no electron-hole pair present, the single exciton ground state, (labelled as $|2\rangle$), and the first excited state of the exciton (labelled as $|1\rangle$). The qubit is based on the two level system formed by $|0\rangle$ and $|1\rangle$. The exciton ground state $|2\rangle$ is a spectator state used to monitor the population of state $|1\rangle$. This is possible because $|1\rangle$ decays non-radiatively to $|2\rangle$ long before it can radiatively decay to $|0\rangle$. The state $|1\rangle$ then decays radiatively to $|0\rangle$ and is detected as the photoluminescence (PL) signal as summarized in Fig. 1(a). Such a detection scheme has been described in [11] and [12].

The qubit is manipulated by a pulsed Ti:Sapphire laser (in resonance with the $|0\rangle \rightarrow |1\rangle$ transition), with the pulse width adjustable from 4 to 10 ps and with a repetition rate of 80 MHz. The photoluminescence ($|2\rangle \rightarrow |0\rangle$ transition) is collected along the normal direction to the sample (maintained at 5 K), dispersed by a spectrograph and imaged using a liquid nitrogen cooled two-dimensional array detector. Shown in Fig. 1(b) is the spectral image of the particular QD used for this study on an unprocessed sample. Light with linear polarization was used in such a way as to excite only one transition in the fine-structure split doublet [13].

Figure 1(c) shows the PL intensity as a function of the square root of the average laser intensity. Note that in each data series, the laser pulse width is fixed while the intensity is varied over several orders of magnitude to scan the input pulse area. The oscillations of the PL intensity correspond to the Rabi rotations described above that the population of state $|1\rangle$ undergoes as a function of the input pulse area, $\theta = \frac{\mu}{\hbar} \int\limits_{-\infty}^{\infty} \varepsilon(t')dt'$ [14].

Here $\mu$ is the transition dipole moment and $\varepsilon(t)$ the electric field envelope. Since the laser pulse shape is known, $\int \varepsilon(t')dt'$ for each pulse can be calculated. Thus, from the periodicity one can find the transition dipole moment, $\mu = 40$ Debye. Furthermore, the variation of the oscillation periodicity in the three oscillations of Fig. 1(c) agrees quantitatively with the fact that the input pulse area should be proportional to the square root of the pulse width at the same average laser power. One also observes that at a fixed pulse width $\tau_p$ the RO amplitude is damped out as $\theta$ increases. This is emphasized in Fig. 1(d) where we plot the amplitudes extracted from the oscillations in Fig. 1(c) at $\theta = \pi$, $2\pi$, etc. on a logarithmic scale. In addition, the smaller the pulse width $\tau_p$, the faster the RO is damped out when the laser intensity is increased. However, at $\theta < 2\pi$, the smaller pulse width actually results in larger RO amplitude. We believe that the dependence of the damping rate on the pulse width (which is proportional to the inverse spectral width) is important to understand its origin.

One fundamental question arises: "Does the extra damping occur only during the manipulation pulse or does it persist even after the pulse is over?" To answer this question we performed wavepacket interferometry experiments under different excitation intensities on the same QD. The quantum interference amplitude is measured while varying the time delay between the pulses for a given single pulse input area. This measurement probes the decoherence rate in the time interval *between* two laser pulses. A detailed description of this procedure is given elsewhere for both linear [3] and nonlinear excitation regimes [6,7]. We find that $T_2$ decreases by a factor of two from $\theta << \pi$ to $\theta = 2.5\pi$ (Fig. 2). This shows that the RO damping of Fig. 1(c) indeed originates in an excitation dependent dephasing term. Moreover, this perturbation persists even after the end of the pulse.

In order to fully capture the dynamics of RO damping, the excitonic state is described by a three-level system where the laser exclusively interacts with the $|0\rangle \rightarrow |1\rangle$ transition. The middle state $|2\rangle$ acts as a shelving state whose time integrated population $\int_0^\infty \rho_{22} dt$ is proportional to the PL intensity. The density matrix equations read:

$$\frac{d\rho_{11}}{dt} = i\frac{\Omega}{2}(\rho_{10} - \rho_{01}) - (\gamma_1 + \kappa + \zeta)\rho_{11}$$

$$\frac{d\rho_{00}}{dt} = -i\frac{\Omega}{2}(\rho_{10} - \rho_{01}) + \gamma_1\rho_{11} + \gamma_2\rho_{22} + \zeta\rho_{11}$$

$$\frac{d\rho_{01}}{dt} = -i\delta\rho_{01} - i\frac{\Omega}{2}(\rho_{11} - \rho_{00}) - (\frac{\gamma_1 + \kappa + \zeta + 2\gamma^*}{2})\rho_{01}$$

$$\frac{d\rho_{22}}{dt} = -\gamma_2\rho_{22} + \kappa\rho_{11}$$

where $\Omega = \Omega(t) = \mu\varepsilon(t)/\hbar$ is the Rabi frequency, $\delta = \omega_0 - \omega$ is the detuning from the resonance frequency $\omega_0$ of the $|0\rangle$ to $|1\rangle$ transition ($\omega$ is the laser frequency) and $\gamma_{1,2}, \kappa, \zeta, \gamma^*$ are damping terms whose effect is depicted in the energy diagram of Fig. 1(a). $\gamma_1$ and $\gamma_2$ denote the radiative recombination rates of state $|1\rangle$ and $|2\rangle$, respectively. $\kappa$ is the decay rate from state $|1\rangle$ to state $|2\rangle$ which primarily occurs via phonon emission [12]. $\gamma^*$ describes pure dephasing (dephasing without population relaxation) and $\zeta$ is an additional decay rate from state $|1\rangle$ to state $|0\rangle$ that accounts for all other processes that scatter the exciton out of state $|1\rangle$ without decaying into $|2\rangle$. Radiative lifetimes in our sample are generally larger than 500 ps so that $\gamma_1, \gamma_2 << 1/\tau_p$ and thus they play no significant role in the dephasing process. The overall dephasing rate of $|1\rangle$ then becomes:

$$\frac{1}{T_2} = \frac{\kappa + \varsigma}{2} + \gamma^*$$

From numerical integration of the density matrix equations at exact resonance ($\delta = 0$) and with the initial conditions $\rho_{11} = \rho_{22} = 0$ and $\rho_{00} = 1$ one finds that there can be no decay of the RO amplitude with intensity unless the excited state dephasing rate increases with intensity. Throughout we assume that at low intensity, $\kappa = 2/T_2^{(0)}$ where $T_2^{(0)}$ is obtained from wavepacket-interferometry.

In principle, all three parameters, $\kappa$, $\varsigma$, and $\gamma^*$ can depend on the excitation intensity and result in intensity dependent damping of ROs. However, each affects the damping behaviour in a different way (Fig. 3). For example, one can choose intensity dependent $\varsigma$ that gives rise to correct damping of the ROs, however, the background of the oscillations also decreases (double-dashed curve in Fig. 3). This is not surprising because $\varsigma$ corresponds to scattering of the exciton out of the QD and will not contribute to the PL signal. The oscillations are also asymmetric if the phonon decay rate $\kappa$ is intensity dependent (dashed curve in Fig. 3). On the other hand, pure dephasing ($\gamma^*$ term) only damps out the coherence without eliminating the excitonic state, thus resulting in damped oscillations about the center line (solid curve in Fig. 3). We have found that all three curves in Fig. 1(c) can be fitted with a pure dephasing term of the form $\gamma^* = c \cdot I/\tau_p \propto \Delta\lambda \cdot I$ where $\Delta\lambda \propto 1/\tau_p$ is the laser bandwidth, $I$ is the average laser intensity, and $c$=0.4 mW$^{-1}$ [solid lines in Fig. 1(c)]. While we cannot completely exclude a more complicated relationship between $\gamma^*$ and $I$ and/or a combination of intensity dependent parameters we believe the above choice to be most reasonable. Thus we conclude that the RO damping during active manipulation is primarily due to the additional pure dephasing term induced at high excitation intensity. This behaviour also rules out the mechanism resulting from coupling to delocalized excitons, proposed in [5] for interface fluctuation QDs (IFQDs) since that mechanism will take the excitonic state out of the QDs and will give rise to totally different overall behaviour. This is not surprising since the energy confinement in SAQDs is much higher than that in IFQDs. We note that although pure dephasing does not play an important role in IFQDs [15,16], its manifestation has been reported in SAQDs [17].

What could be the underlying mechanism? The lattice mediated dephasing model proposed in [18], showed that the RO amplitudes decrease with the laser intensity. However, the pulse width dependence is inconsistent with our experimental observation. Bi-excitonic excitation is another possibility since the Rabi energy $\hbar\Omega$ in our experiments could be close to the bi-exciton binding energy (typically a few meV). However, experiments performed using circularly polarized light to suppress bi-exciton excitation showed almost identical intensity dependent behaviour, thus ruling out this possibility. We note, nevertheless, that bi-excitonic scattering could contribute when shorter pulses are used such as in [19] and has been investigated theoretically [20]. Inter-dot localized-exciton interactions (dipole-dipole) were also considered. However, theoretical calculations [16] showed that the interaction energy is only a few $\mu$eV at a typical inter-dot distance, too small to give rise to significant damping.

Below we provide strong evidence that the observed RO damping in our system is due to indirect excitation of carriers in the wetting layer (WL) that has compositional fluctuation. Recent work [21] has clarified the origin of the continuous absorption background related to the wetting layer and attributed the broad resonances seen in single dot PL excitation (PLE) spectra [12, 22] to bound-to-continuum and continuum-to-bound transitions. Such indirect excitation channels involving a hole (electron) in the WL and an electron (hole) in other QDs can exist, despite weak transition dipole moments (the wavefunctions of the electronic states in the QDs decay rapidly into the WL) [Fig. 4(a)]. Since ROs are excited under very strong excitation, these low probability channels can be excited. Moreover, since the WL has compositional fluctuations [23] the phase space for such transitions is large. Once the carriers in the WL are excited, they provide a dephasing channel for the excitonic states in the QD that exhibit ROs. The linear dependence on the intensity for RO damping (one carrier type is sufficient to cause dephasing) and their behaviour with the pulse width, i.e. spectral width, is consistent with coupling to a continuum of states.

In order to verify that such processes indeed occur, we probed QDs within a submicron shadow mask under varying excitation conditions. In this case, at most about ~150 QDs can be excited so that resonant and above-band PL spectra can be conveniently compared. Figure 4(b) shows part of the PL spectrum under resonant excitation (~1.33 eV) for three increasing intensities (solid lines) $I_0$=0.06 mW, $6I_0$, and $23I_0$, from top to bottom, respectively. At low power, only QDs with their excited states in resonance with the laser emit. Most other QDs have energy states far from resonance and cannot be excited. Similarly, direct excitation of excitons in the WL is not possible because the absorption edge is far above the laser frequency. However, when the laser intensity is increased, transitions involving the excitation of one electron in the WL and one hole in the off-resonance QD, or vice versa, can occur, albeit their weak oscillator strengths. This allows to populate the off-resonance QDs and the wetting layer, as is evident by the fact that at higher intensity, more off-resonance QDs emit, and eventually the emission spectrum becomes very similar to the spectrum excited above the band edge [dashed curve in Fig. 4(b)]. Thus, the absorption spectrum of a single QD is highly dependent upon excitation power. As is shown in Fig. 4(c), only the peaks present at low intensity are truly excited coherently. For instance, the QD state labelled by QD#1 undergoes RO while the PL from QD labelled QD#2 increases with the square root of the intensity. In contrast, the PL from another QD (labelled QD#3) and the background "wetting layer" emission increases superlinearly with intensity. More interestingly, such a PL displays a very similar pulse width, i.e. spectral width dependence as the RO damping rate [Fig. 1(d)]. This is consistent with multi-event processes ($I^n$ intensity dependence) involving a continuum as described in [21]. The carriers thereby created in the WL interact with the exciton undergoing RO leading to intensity dependent dephasing.

Finally we note that we also considered processes such as two-photon absorption, presumed to be responsible for the up-converted PL at the band-edge under strong excitation [24]. Theoretical estimates of the resulting scattering rate for an exciton in an excited state come short by several orders of magnitude and would not significantly affect the dephasing process.

In summary, we have provided strong evidence that the decoherence processes during strong field manipulation of excitonic qubits in SAQDs are primarily due to the indirect excitation of carriers in the wetting layer. If one can suppress the compositional fluctuation of the WL, then the major source of decoherence will also be suppressed, thus raising the quality factor of qubits to a practical regime.

We thank Professor L.J. Sham for fruitful discussions. This work was supported by NSF-NIRT (DMR-0210383), NSF-FRG (DMR-0071893 and DMR-0306239), Texas Advanced Technology program, and the W.M. Keck Foundation.

**(References and Notes)**

**(figure captions)**

FIG. 1. Rabi oscillations of the upper state in the excitonic three-state system and its PL detection. (a) QD energy diagram. The QD is resonantly excited to the first excited excitonic state $|1\rangle$. The population that relaxes non-radiatively to the excitonic ground state $|2\rangle$ is eventually emitted and detected as the PL signal. The different decay channels and their rate constants are denoted by arrows. (b) Spectral image of QDs excited at 1.3418 eV. The QD investigated is marked by an arrow and is well isolated, both spatially and spectrally. The total vertical dimension is about 10 μm. (c) Rabi oscillations for different pulse widths. The PL from the $|2\rangle$ to $|0\rangle$ transition was recorded while the average intensity was varied for fixed pulse width $\tau_p$. The three curves have been displaced for clarity. The fit (solid lines) was obtained by numerical integration of the density matrix equations using a pure dephasing term proportional to the intensity. Note that $\theta$ is proportional to $E_0\tau_p$, whereas the average intensity is proportional to $E_0^2\tau_p$, where $E_0$ is the peak electric field amplitude. The oscillations are therefore periodic in the square root of the average intensity and the period scales with $\tau_p^{-1/2}$. (d) Negative logarithm of the oscillation amplitude plotted versus the input pulse area. The data points are taken from the peaks and valleys of the ROs shown in (c), corresponding to the points where $\theta = n\pi$. Note that the longer the pulses, the weaker the damping. The fitted lines are a guide to the eye.

FIG. 2. Quantum interference amplitude (logarithmic scale) for different coarse time delays under low ($\theta \ll \pi$, open squares) and high ($\theta = 2.5\pi$, filled squares) excitation intensity. Indicated are the dephasing times $T_2$ obtained from the linear fits.

FIG. 3. Simulated ROs assuming various intensity dependent decay terms. Plotted is the case when either $\kappa = \kappa_0 + c' \cdot I / \tau_p$ (dashed curve), $\varsigma = c \cdot I / \tau_p$ (double-dashed curve), or $\gamma^* = c \cdot I / \tau_p$ (solid curve). The data for $\tau_p = 7.0$ ps is plotted as a reference (squares).

FIG. 4. PL spectra of QDs under a 1 micron aperture and their power dependence. (a) Band diagram along a direction perpendicular to the growth direction. The dark bands between dots represent a continuum of states resulting from a fluctuating wetting layer. The dashed arrows indicate the transitions that can occur at high intensity and are likely responsible for the superlinear dependence of the background signal. (b) resonant and non-resonant PL spectra. The dots were excited resonantly at ~1340 meV at intensities $I_0$=0.06 mW, $6I_0$ and $23I_0$, top to bottom, respectively (solid lines) and above band at ~1650 meV (dashed line). (c) Intensity dependence of peaks denoted by QD#1, QD#2, QD#3 in (a) and the background signal. Note that the PL of peak QD#1 is plotted versus the input pulse area. For peak QD#3 and the background signal, for which the PL grows superlinearly with intensity, the experiment was repeated for three laser pulse widths $\tau_p =$ 4.5 ps (squares), $\tau_p = 5.5$ ps (triangles) and $\tau_p = 7.6$ ps (diamonds). The smaller the pulse width (the larger the spectral width), the stronger the PL intensity.

**(figures)**

FIG. 1.

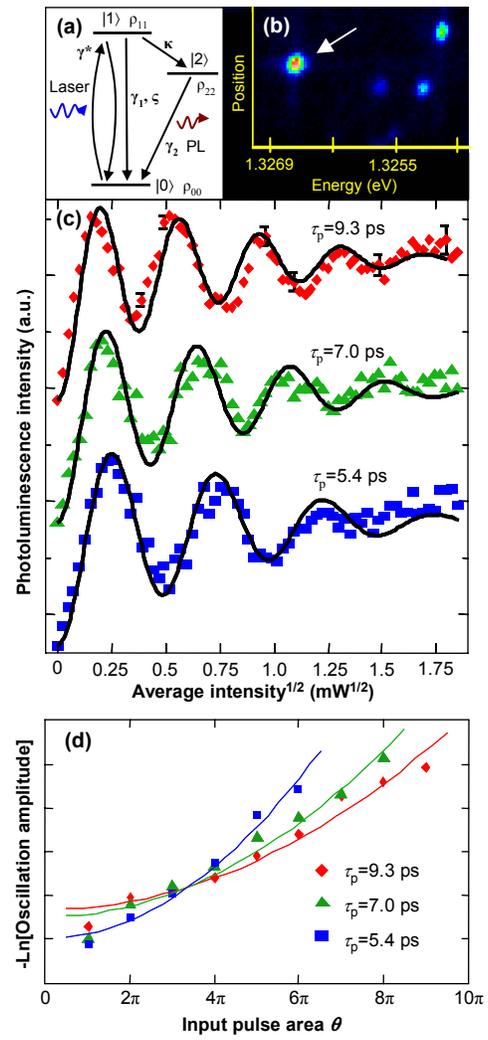

FIG. 2.

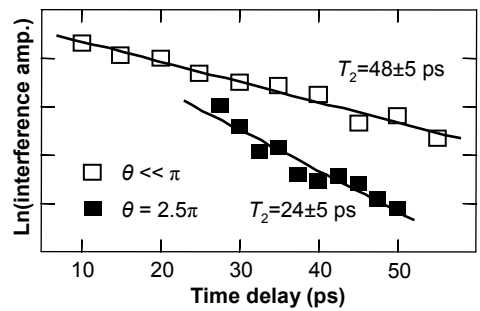

FIG. 3.

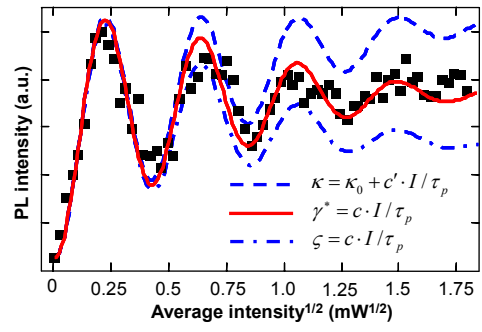

FIG. 4.

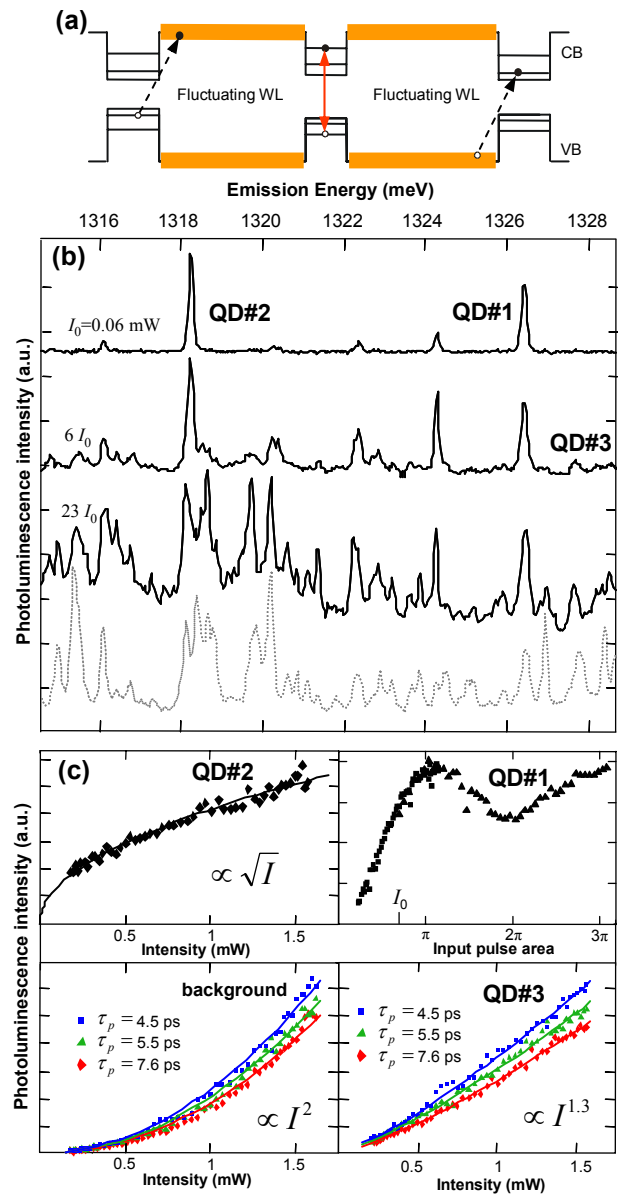